\documentclass{aa} 

\def\dosingle#1::::{#1}  \def\dodouble#1::::{ } 

\dodouble \documentclass[referee]{l-aa} ::::

\usepackage{natbib}

\def\nice#1::::{#1}    \def\subm#1::::{}   

\newcommand\zzz[2]{#2}  

\def\SS{Sect.~}

\def\apj{ApJ}                 
\def\aj{AJ}                       
\def\aap{A\&A}            
\def\aaps{A\&AS}            
\def\mnras{MNRAS}
\def\araa{ARA\&A}

\nice \input epsf ::::

\renewcommand\citep[1]{(\citealt{#1})}
\newcommand\citepf[1]{(\citealt*{#1})}    

\def\centreline{\centerline}
\def\.{{\cdot}} 
\def\gtapprox{\,\lower.6ex\hbox{$\buildrel >\over \sim$} \, }
\def\ltapprox{\,\lower.6ex\hbox{$\buildrel <\over \sim$} \, }
\def\propapprox{\,\lower.6ex\hbox{$\buildrel \propto\over \sim$} \, }

\def\arcs{\ifmmode {'' }\else $'' $\fi}     
\def\arcm{\ifmmode {' }\else $' $\fi}       
\def\deg{\ifmmode^\circ\else$^\circ$\fi}    

\def\fr7{7$ \hskip -0.9ex \vrule height0.8ex width0.8ex depth-0.73ex
                                                                \hskip0.1ex$}
\def\frtoday{Le\space\number\day\space\ifcase\month\or
  janvier\or f\'evrier\or mars\or avril\or mai\or juin\or
  juillet\or ao\^ut\or septembre\or octobre\or novembre\or d\'ecembre\fi\space \number\year}

\newcommand\joref[5]{#1, #5, {#2 }{#3, } #4}  
 
\def\cqg{ClassQuantGra}   %

\def\hMpc{\mbox{$\,h^{-1}$ Mpc}}
\def\hMpcinv{\mbox{$\,h\,$Mpc$^{-1}$}}

\def\llss{L_{\mbox{\rm \small LSS}}}
\def\Ntot{N_{\mbox{\rm \small tot}}}
\def\pgauss{P_{\mbox{\rm \small gauss}}}
\def\CIV{C~{\sc IV}}
\def\CIII{C~{\sc III}}
\def\MgII{Mg~{\sc II}}
\def\numax{\nu_{\mbox{\rm \small max}}}
\def\pgauss{P_{\mbox{\rm \small gauss}}}
\def\sqdeg{~sq.~deg.}
\def\dpm{d_{\mbox{\rm \small pm}}}
\def\dperp{d_\perp}

\begin{document}

\thesaurus{2 
(12.03.3; 
12.03.4; 
11.17.3; 
12.04.3; 
12.12.1; 
05.18.1) 
}

\title{Tangential Large Scale Structure as a Standard Ruler: Curvature Parameters from Quasars}

\author{B. F. Roukema\inst{1}  
\and
 G.~A. Mamon\inst{2,3}}

\authorrunning{B. F. Roukema \& G.~A. Mamon}
\titlerunning{Curvature Parameters from QSO Tangential LSS}

   \offprints{B. F. Roukema}

 \institute{Inter-University Centre for Astronomy and Astrophysics, 
    Post Bag 4, Ganeshkhind, Pune, 411 007, India {\em (boud@iucaa.ernet.in)}
\and
Institut d'Astrophysique de Paris
(CNRS UPR 341),
98bis Bd Arago, F-75014 Paris, France 
{\em (gam@iap.fr)}
\and
DAEC (CNRS UMR 8631),
Observatoire de Paris-Meudon, 5 place Jules Janssen, F-92195 Meudon Cedex,
France}

\date{\frtoday}

\maketitle

\begin{abstract}
Several observational analyses suggest that matter is spatially
structured at a scale of 
$\llss \approx 130${\hMpc} at 
low redshifts. This peak in the power spectrum provides 
{\em a standard
ruler in comoving space} 
which can be used to compare the local geometry at high and 
low redshifts, thereby constraining the curvature parameters.

It is shown here that this power spectrum peak is present 
in the observed quasar distribution at $z\sim 2$: 
qualitatively, via wedge diagrams which clearly
show a void-like structure, and quantitatively, 
via one-dimensional Fourier analysis of 
the quasars' tangential distribution. The sample studied here
contains 812 quasars.

The method produces strong constraints 
(68\% confidence limits)
on the density parameter $\Omega_0$
and weaker constraints on the cosmological constant $\lambda_0$, which
can be expressed by the relation 
$ \Omega_0 =  (0.24 \pm0.15) + (0.10\pm0.08)\,\lambda_0 $.
Independently of $\lambda_0$ (in the range $\lambda_0 \in [0,1]$), 
the constraint is
$0.1 < \Omega_0 < 0.45$.
Constraints if the cosmological constant is zero or
if $\lambda_0 \equiv 1-\Omega_0$ are 
$\Omega_0=0.24^{+0.05}_{-0.15}$ and 
$\Omega_0=0.30 \pm 0.15$ respectively.

The power spectrum peak method is independent from  
the supernovae Type Ia method by choice of astrophysical object, 
by redshift range, and by use of a standard ruler instead of 
a standard candle. Combination of the two results yields
$\Omega_0 =  (0.30\pm0.11) + (0.57\pm0.11) (\lambda_0-0.7),$  
$0.55 < \lambda_0 < 0.95,$ (68\% confidence limits)   
{\em without assuming that 
$\lambda_0\equiv 1-\Omega_0$.} This strongly supports the possibility that 
the observable universe satisfies a nearly flat, perturbed
Friedmann-Lema\^{\i}tre-Robertson-Walker model, {\em independently
of any cosmic microwave background observations}.

\end{abstract}


\dodouble \clearpage :::: 


\def\tfourier{
\begin{table}
\caption{Definition of the two tangentially 
long, densely observed,
homogeneous, subsamples of 
the \protect\citet{IovCS96} quasar survey listed 
in the \protect\citet{Veron98} catalogue, 
near the south galactic pole (SGP). The right ascension (first row)
and declination (second row) subsamples are defined by 
J2000 limits 
($\alpha_1 \le \alpha \le \alpha_2$, 
$\delta_1 \le \delta \le  \delta_2$). For the purpose of 
Poisson simulations, these are subdivided in 
right ascension at ($\alpha', \alpha'' $) and in
declination at ($\delta', \delta''$), in order to 
allow for the possibility of 
different magnitude zero points or different magnitude cutoffs in 
the different plates. The number of objects $N$ in each subsample 
is indicated. The total number of physically distinct quasars 
in the two subsamples is $\Ntot= 812$.
\label{t-fourier}}
$$\begin{array}{cccc cc cc c} \hline 
\alpha_1 & \alpha_2 & \delta_1 & \delta_2 &  \alpha' & \alpha'' 
& \delta' & \delta'' & N\\ 
\hline
\multicolumn{9}{l}{\mbox{\rm `Right ascension ($\alpha$) subsample'}} \\
0^h42^m & 1^h59^m &   -42.0 & -37.5 & 1^h07^m & 1^h33^m & && 604\\ 
\multicolumn{9}{l}{\mbox{\rm `Declination ($\delta$) subsample'}} \\
0^h42^m & 1^h00^m &   -42.0 & -28.0 & & & -37.5 & -32.5 & 373 \\
\hline
\end{array}$$
\end{table}
}  

\def\tborders{
\begin{table}
\caption{Angular separation 
of plate centres (table~1, \protect\citealt{IovCS96}) 
from the four borders [$\alpha$(1),(2), $\delta$(1),(2) in table, 
from the borders in strictly increasing numerical $\alpha$ or $\delta$
values] 
and maximum centre-corner angle (3) of the fields 
for the right ascension ($\alpha$) and declination ($\delta$) subsamples, 
defined here 
in Table~\protect\ref{t-fourier}, in great circle degrees. 
Field numbers (\#) are ESO/SERC field numbers.
\label{t-borders}}
$$\begin{array}{l ccc ccc } \hline 
    & \multicolumn{3}{c}{\alpha} &\multicolumn{3}{c}{\delta}\\
\hline
\#: & 295 & 296 & 297 & 295 & 351 & 411 \\

\alpha(1)&    2.37 &     2.55 &     2.53 &     2.37 &     1.73 &     1.40 \\
\alpha(2)&    2.43 &     2.45 &     2.47 &     1.09 &     1.97 &     2.51 \\
\delta(1)&    2.27 &     2.26 &     2.25 &     2.27 &     2.77 &     2.77 \\
\delta(2)&    2.23 &     2.24 &     2.25 &     2.23 &     2.23 &     1.73 \\
(3)&    3.33 &     3.41 &     3.38 &     3.29 &     3.40 &     3.74 \\
\hline
\end{array}$$
\end{table}
}  

\def\fzhist{
\begin{figure}
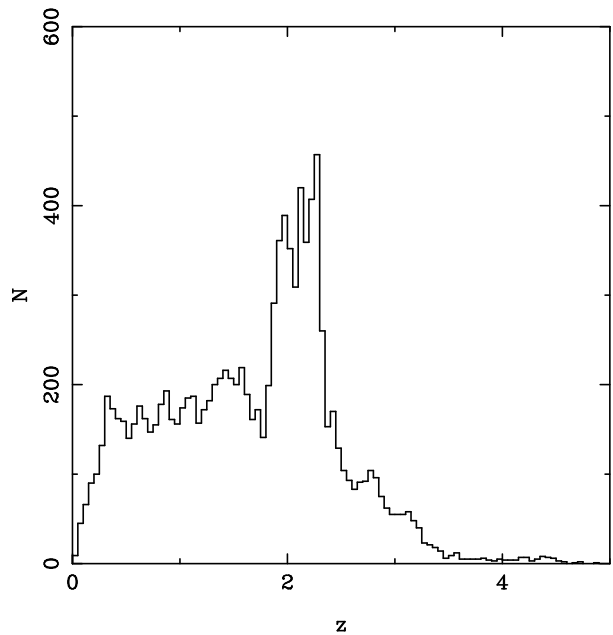

\centering 
\nice \centreline{\epsfxsize=8cm
\zzz{\epsfbox[56 43 459 463]{"`gunzip -c zhist.ps.gz"} }
{\epsfbox[56 43 459 463]{"zhist.ps"}}  } ::::

\caption[]{ 
Redshift distribution of quasars in the compilation of
\protect\citet{Veron98}.
}
\label{f-zhist}
\end{figure}
} 

\def\fallsky{
\begin{figure}
\centering 
\nice \centreline{\epsfxsize=9.5cm
\zzz{\epsfbox[16 22 784 596]{"`gunzip -c allsky.ps.gz"} }
{\epsfbox[16 22 784 596]{"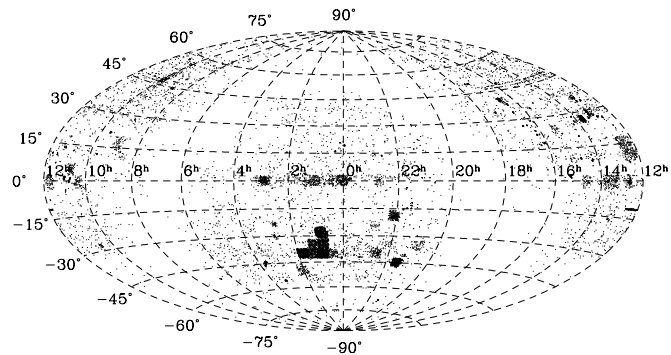"}}  } ::::
\caption[]{ 
Sky distribution of quasars in the compilation of
\protect\citet{Veron98}. 
}
\label{f-allsky}
\end{figure}
} 

\def\fzVeron{
\begin{figure}
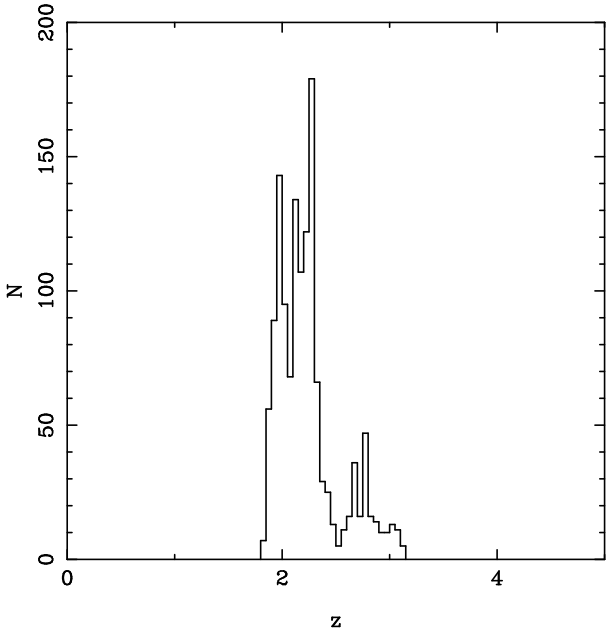

\centering 
\nice \centreline{\epsfxsize=8cm
\zzz{\epsfbox[56 43 459 452]{"`gunzip -c zVeron.ps.gz"} }
{\epsfbox[56 43 459 452]{"zVeron.ps"}}  } ::::
\caption[]{ 
Redshift distribution of quasars with redshifts from
\protect\citet{IovCS96}.
}
\label{f-zVeron}
\end{figure}
} 

\def\fskyV{
\begin{figure}
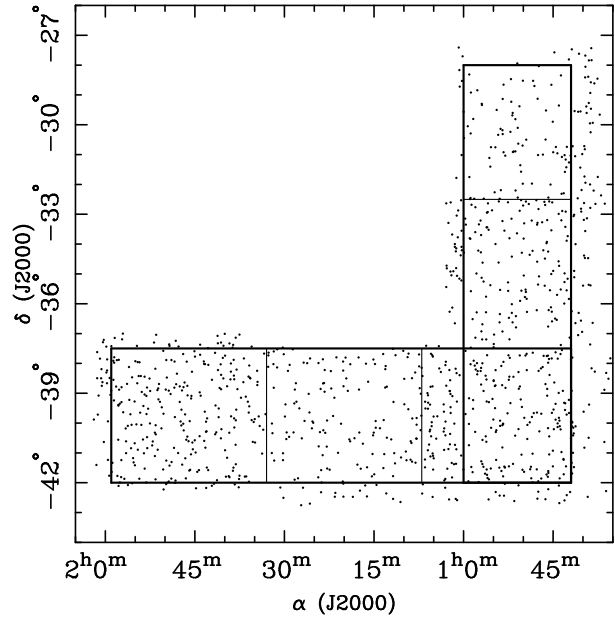

\centering 
\nice \centreline{\epsfxsize=8cm
\zzz{\epsfbox[56 43 459 452]{"`gunzip -c skyV.ps.gz"} }
{\epsfbox[56 43 459 452]{"skyV.ps"}}  } ::::
\caption[]{ 
Sky distribution of quasars with redshifts from
\protect\citet{IovCS96}, which lie in the redshift
range $1.8 \le z < 2.4$. 
Thick lines show the outlines of the right ascension and 
declination samples chosen for one-dimensional Fourier analysis. 
Thin vertical and horizontal lines 
show divisions in right ascension and declination 
in the right ascension and declination subsamples respectively, 
which simulated distributions are Poisson distributed. 
See Table~\protect\ref{t-fourier} for numerical values of these
limits.
}
\label{f-skyV}
\end{figure}
} 

\def\fhyponera{
\begin{figure}
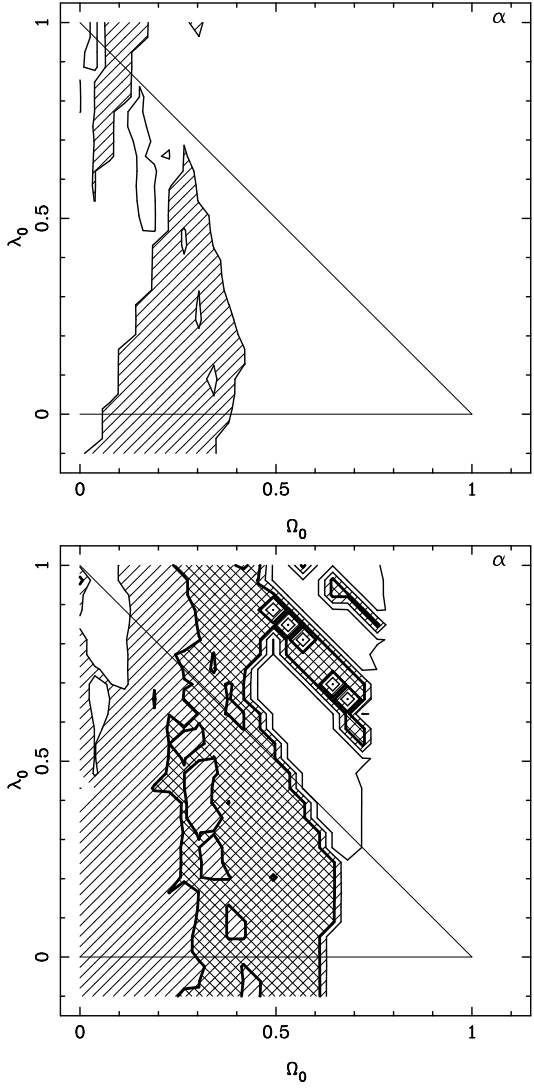

\centering 
\nice \centreline{\epsfxsize=7cm
\zzz{\epsfbox[56 38 459 452]{"`gunzip -c hyp1ra_A.ps.gz"} }
{\epsfbox[56 38 459 452]{"hyp1ra_A.ps"}}  } ::::
\nice \centreline{\epsfxsize=7cm
\zzz{\epsfbox[56 38 459 452]{"`gunzip -c hyp1ra_f.ps.gz"} }
{\epsfbox[56 38 459 452]{"hyp1ra_f.ps"}}  } ::::

\caption[]{ 
Confidence levels for the different
criteria for trying to reject ${\cal H}_1$, for the right
ascension subsample.
The upper panel is for $f(\nu_0)$, the lower panel for $\numax$.
Rejection of the null hypothesis ${\cal H}_1$ at
$> 0\,\sigma,$ $> 1\,\sigma,$ $>2\,\sigma$ and $> 3\,\sigma$ confidence levels, i.e. 
$1-P > 50\%,$ $1-P > 84\%,$ $1-P > 98\%$ and $1-P > 99.9\%$,
is shown by contoured regions with light shading, light cross-hatched,
medium cross-hatched and heavy cross-hatched shading, respectively.
(In a few of this set of figures, a few contours at
$< 0\,\sigma$ and $< -1\,\sigma$ are also shown. These are not useful
for null hypothesis rejection.)
Lines indicating 
$\lambda_0 =0$ and $\lambda_0=1-\Omega_0$ are shown for $\Omega_0 \le1$
as a guide to the eye.
}
\label{f-hyp1ra_}
\end{figure}
} 

\def\fhyponedec{
\begin{figure}
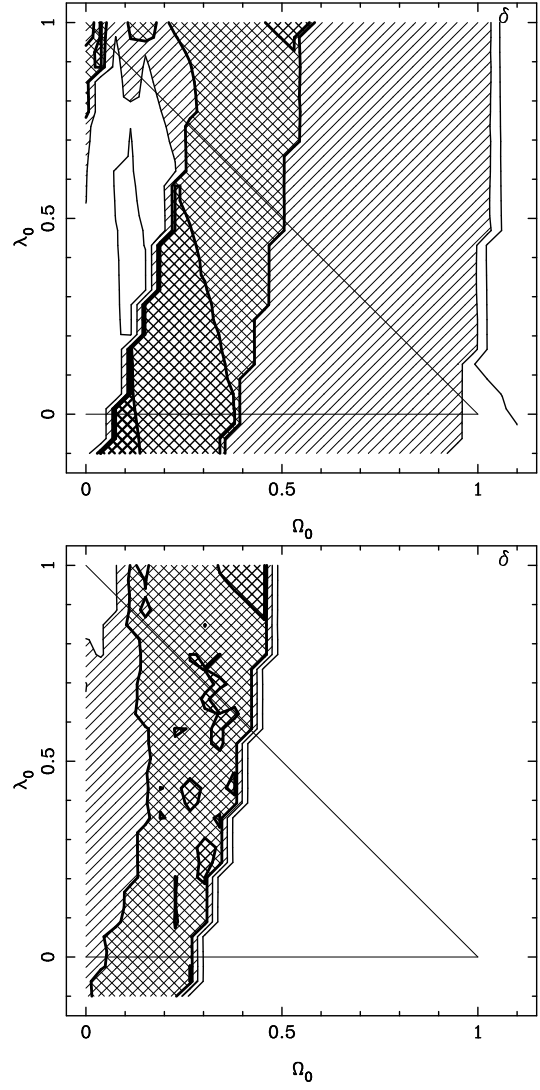

\centering 
\nice \centreline{\epsfxsize=7cm
\zzz{\epsfbox[56 38 459 452]{"`gunzip -c hyp1decA.ps.gz"} }
{\epsfbox[56 38 459 452]{"hyp1decA.ps"}}  } ::::
\nice \centreline{\epsfxsize=7cm
\zzz{\epsfbox[56 38 459 452]{"`gunzip -c hyp1decf.ps.gz"} }
{\epsfbox[56 38 459 452]{"hyp1decf.ps"}}  } ::::

\caption[]{ 
Confidence levels for rejecting ${\cal H}_1$, for the 
declination subsample.
The upper panel is for $f(\nu_0)$, the lower panel for $\numax$.
Shading is as for Fig.~\protect\ref{f-hyp1ra_}.

}
\label{f-hyp1dec}
\end{figure}
} 

\def\fhyponeall{
\begin{figure}
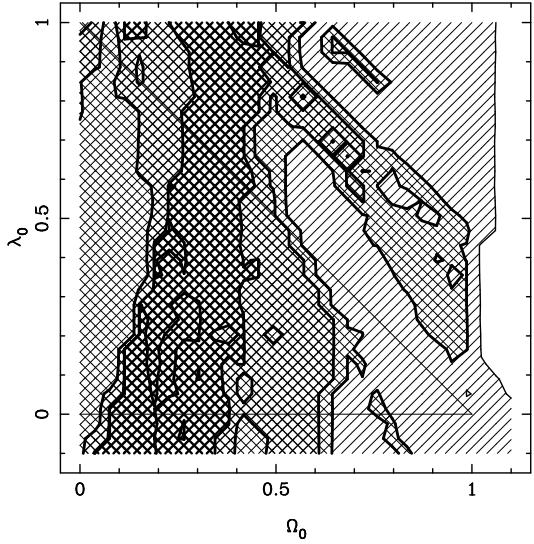

\centering 
\nice \centreline{\epsfxsize=7cm
\zzz{\epsfbox[56 38 459 452]{"`gunzip -c hyp1tldx.ps.gz"} }
{\epsfbox[56 38 459 452]{"hyp1tldx.ps"}}  } ::::

\caption[]{ 
Combined confidence levels for rejecting ${\cal H}_1$, 
using Eq.~(\protect\ref{e-confall}) to combine the results
shown in Fig.~\protect\ref{f-hyp1ra_} 
and Fig.~\protect\ref{f-hyp1dec}.
Shading is as for Fig.~\protect\ref{f-hyp1ra_}.

}
\label{f-hyp1all}
\end{figure}
} 

\def\fhyptwora{
\begin{figure}
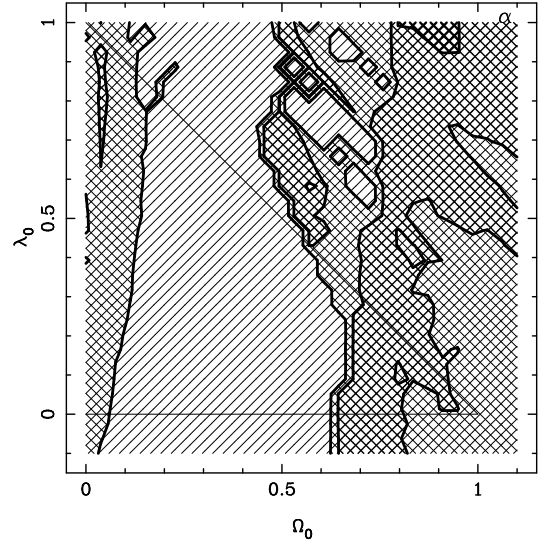

\centering 
\nice \centreline{\epsfxsize=7cm
\zzz{\epsfbox[56 38 459 452]{"`gunzip -c hyp2ra_f.ps.gz"} }
{\epsfbox[56 38 459 452]{"hyp2ra_f.ps"}}  } ::::

\caption[]{ 
Confidence intervals for rejecting 
${\cal H}_2$, the hypothesis that the large scale structure peak
occurs at $1/\llss$,
for the right ascension subsample. Shading styles
are as for the previous figures, except that the confidence
levels are two-sided, i.e. the four successively darker 
shadings are
for 
$1-P > 0\%,$ $1-P > 68\%,$ $1-P > 95\%$ and $1-P > 99.7\%$
respectively. 
}
\label{f-hyp2ra_}
\end{figure}
} 

\def\fhyptwodec{
\begin{figure}
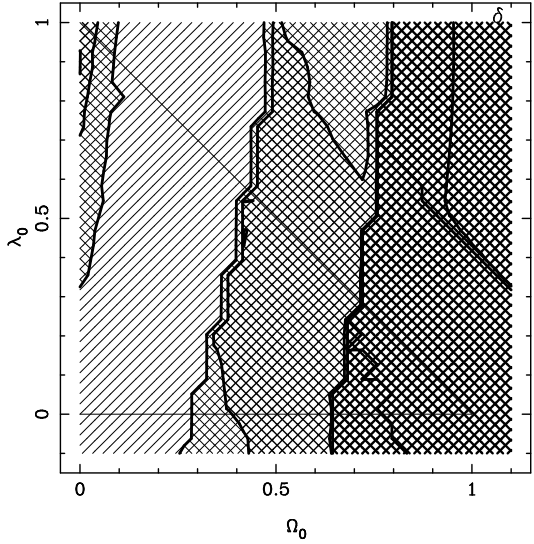

\centering 
\nice \centreline{\epsfxsize=7cm
\zzz{\epsfbox[56 38 459 452]{"`gunzip -c hyp2decf.ps.gz"} }
{\epsfbox[56 38 459 452]{"hyp2decf.ps"}}  } ::::

\caption[]{ 
Confidence intervals for ${\cal H}_2$, for the declination subsample,
shading as for Fig.~\protect\ref{f-hyp2ra_}.
}
\label{f-hyp2dec}
\end{figure}
} 

\def\fhyptwoall{
\begin{figure}
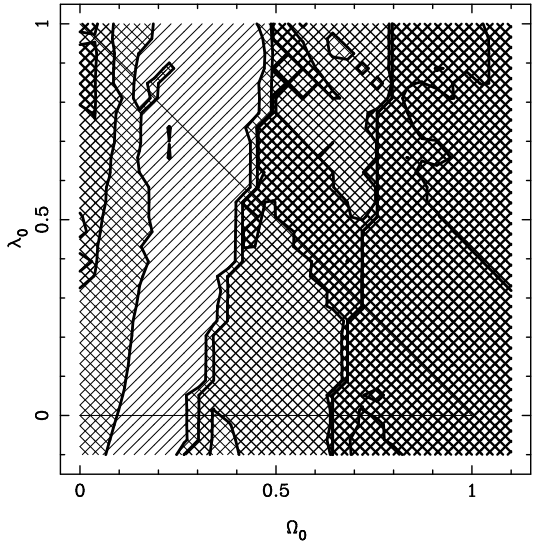

\centering 
\nice \centreline{\epsfxsize=7cm
\zzz{\epsfbox[56 38 459 452]{"`gunzip -c hyp2tldf.ps.gz"} }
{\epsfbox[56 38 459 452]{"hyp2tldf.ps"}}  } ::::

\caption[]{ 
Confidence intervals for ${\cal H}_2$, combining the information
from the two subsamples. 
Shading is as for Fig.~\protect\ref{f-hyp2ra_}.
}
\label{f-hyp2all}
\end{figure}
} 

\def\fhypSNe{
\begin{figure}
\centering 
\nice \centreline{\epsfxsize=7cm
\zzz{\epsfbox[56 38 459 452]{"`gunzip -c hyp2SNef.ps.gz"} }
{\epsfbox[56 38 459 452]{"hyp2SNef.ps"}}  } ::::

\caption[]{ 
Confidence intervals from combining Fig.~\protect\ref{f-hyp2all} 
with the relation $0.8\Omega_0 -0.6\lambda_0 = -0.2\pm0.1$ from
\protect\citet{SCP9812}.
Shading is as for Fig.~\protect\ref{f-hyp2ra_}.
}
\label{f-SNe}
\end{figure}
} 

\def\fhyptwotwenty{
\begin{figure}
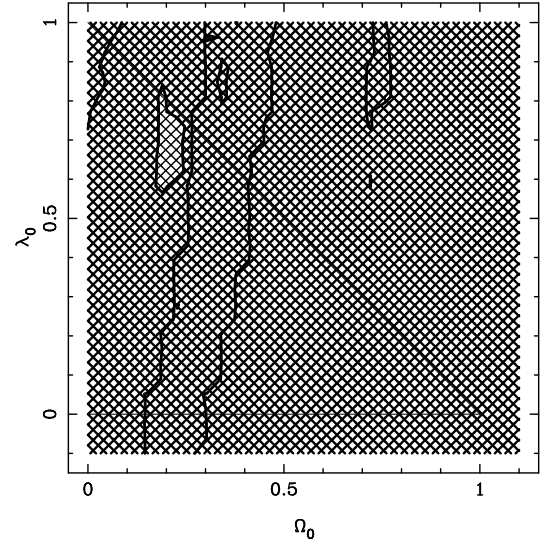

\centering 
\nice \centreline{\epsfxsize=7cm
\zzz{\epsfbox[56 38 459 452]{"`gunzip -c hyp2_20_.ps.gz"} }
{\epsfbox[56 38 459 452]{"hyp2_20_.ps"}}  } ::::
\caption[]{ 
Confidence intervals for ${\cal H}_2$ 
(cf Fig.~\protect\ref{f-hyp2all}), for splitting up both of the
two subsamples into four redshift subsamples each, and increasing the
uncertainty in the power spectrum peak length scale to 
$\Delta \llss=20${\hMpc} (\SS\protect\ref{s-zslices}). 
Shading is as for Fig.~\protect\ref{f-hyp2ra_}, i.e. 
the small patch suggesting a hyperbolic universe model 
is a $1-P > 95\%$ (`$> 2\sigma$') confidence interval region, 
and the rest is for $1-P > 99.7\%$. 
}
\label{f-hyp2_20_}
\end{figure}
} 

\def\fhyptwodfourz{
\begin{figure}
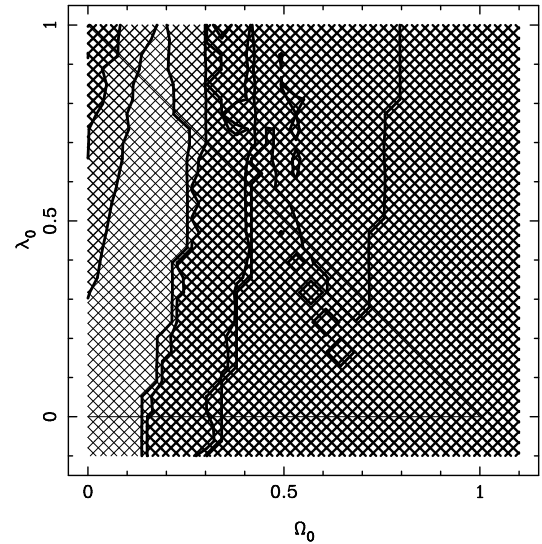

\centering 
\nice \centreline{\epsfxsize=7cm
\zzz{\epsfbox[56 38 459 452]{"`gunzip -c hyp2_d4z.ps.gz"} }
{\epsfbox[56 38 459 452]{"hyp2_d4z.ps"}}  } ::::
\caption[]{ 
Confidence intervals for ${\cal H}_2$ 
(cf Fig.~\protect\ref{f-hyp2dec}), for 
the four redshift subsamples of the declination subsample,
retaining 
$\Delta \llss=10${\hMpc} (\SS\protect\ref{s-zslices}). 
Shading is as for Fig.~\protect\ref{f-hyp2ra_}, i.e. 
the lightest shaded region is for the $1-P > 68\%$ confidence interval.
}
\label{f-hyp2_d4z}
\end{figure}
} 

\def\fdhist{
\begin{figure}
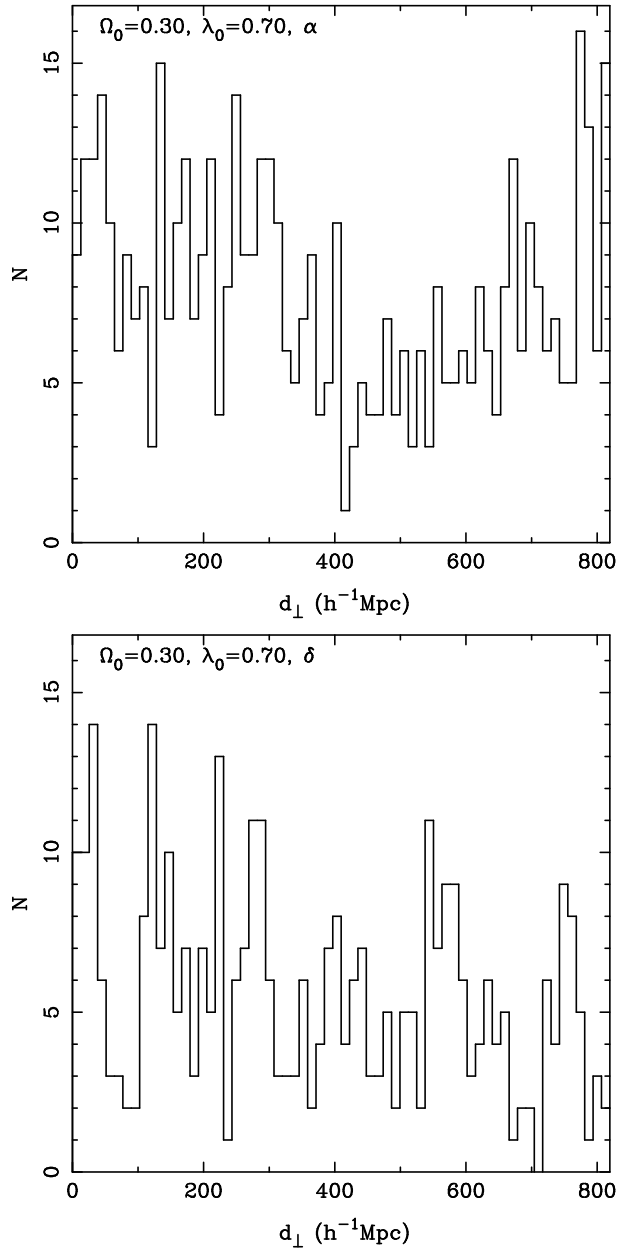

\centering 
\nice \centreline{\epsfxsize=8cm
\zzz{\epsfbox[56 43 459 463]{"`gunzip -c dhista.ps.gz"} }
{\epsfbox[56 43 459 463]{"dhista.ps"}}  } ::::
\nice \centreline{\epsfxsize=8cm
\zzz{\epsfbox[56 43 459 463]{"`gunzip -c dhistd.ps.gz"} }
{\epsfbox[56 43 459 463]{"dhistd.ps"}}  } ::::

\caption[]{ 
Tangential length distribution for $(\Omega_0=0.3, \lambda_0=0.7)$,
for the right ascension (upper panel) and declination 
(lower panel) subsamples (in bins of 12.8{\hMpc}). 
The null hypotheses test relate
to a power spectrum peak at $\llss=130\pm10${\hMpc}.
Differences in number density between
the plate boundaries are not corrected for, these are corrected
for in the simulations.
}
\label{f-dhist}
\end{figure}
} 

\def\fcone{
\begin{figure}
\centering 
\nice \centreline{\epsfxsize=8.5cm
\zzz{\epsfbox[51 40 459 353]{"`gunzip -c conera.ps.gz"} }
{\epsfbox[51 40 459 353]{"conera.ps"}}  } ::::
\nice \centreline{\epsfxsize=8.5cm
\zzz{\epsfbox[51 40 460 336]{"`gunzip -c conedec.ps.gz"} }
{\epsfbox[51 40 460 336]{"conedec.ps"}}  } ::::

\caption[]{ 
Wedge diagrams for the 
for the right ascension (upper panel) and declination 
(lower panel) subsamples,
for $(\Omega_0=0.3, \lambda_0=0.7)$. Since curvature is zero,
rectlinear coordinates are defined: 
$x= \dpm \cos(\theta^* - \theta  ),$ 
$y= \dpm \sin(\theta^* - \theta),$ where
$\theta^* \equiv (\theta_1 + \theta_2)/2 + \pi/2$ and 
$\theta, \theta_1, \theta_2$ and $\dpm$ are defined in 
\SS\protect\ref{s-method1}.
Voids consistent with $\llss=130\pm10${\hMpc} and not due
to selection effects are clearly visible.
}
\label{f-cone}
\end{figure}
} 


\section{Introduction}

It has been known for more than a decade 
(e.g. \citealt*{deLapp86,GH89})
that the spatial distribution of extragalactic objects is structured 
at length scales about an order of magnitude greater than the  
$r_0 \approx 5${\hMpc} scale which characterises galaxy 
clustering via the two-point auto-correlation function.
Observational analyses from several different data sets 
suggest that this is due to a characteristic length scale at
$\llss \approx 130${\hMpc}, or in other words, that 
there is a maximum in the power 
spectrum at $ k = 2\pi/\llss \approx 0.05 ${\hMpcinv} 
(e.g. 
\citealt*{Bro90,Bro99,BJ99,daCosta92,daCosta93,BauE93,BauE94,GazB98,Einasto94,Einasto97corr,Einasto97nat,Deng96} 
or \citealt{Guzzo99} for a recent review).

It has already been suggested that this scale could
be used as a standard ruler which could be compared at low and high
redshifts in order to constrain the curvature parameters, 
$\Omega_0$ (the density parameter) and $\lambda_0$ (the dimensionless
cosmological constant). 
In the redshift direction, at least two analyses 
have been carried out based on this idea:
one analysis of a quasar catalogue 
and one analysis of Lyman break galaxies.

\citet*{Deng94} implicitly used the large scale
structure scale as a curvature constraint in quasar data, 
under the assumption that 
$\lambda_0\equiv 0$, and found that $\Omega_0 \approx 0.4$.
\citeauthor{BJ99} used the radial (redshift)
distribution of Lyman 
break galaxies at $z\sim 3$ 
(table~1, \citealt{GiavDCP98}; fig.~2, \citealt{Adel98}).
They found a correlation
scale of $\Delta z \approx 0.22\pm0.02$, and inferred a relation
$3.2\,\Omega_0 - \lambda_0 \approx 0.7$.

The purpose of this paper is to (i) emphasise that the principle can
be applied to a class of bright objects easily found 
at super-unity redshifts: quasars; (ii) show that redshift
selection effects can be minimised by using the tangential 
distribution instead of the radial distribution; 
and (iii) show pictorially (i.e. qualitatively) 
that quasars do indeed trace large scale structure at $z\sim 2$.

(i) Because quasars are much brighter than Lyman break galaxies,
they offer a potentially much more rapid method of obtaining high
precision estimates of the curvature parameters than the latter. 
Both classes of objects have the advantage relative to supernovae type
Ia \citep{SCP9812,HzS98} of being at super-unity redshifts, so that
the dependence on the curvature parameters is strong.

(ii) In order to avoid the well-known selection effects in the redshift
distribution of quasars (which could also in principle affect the
redshift distribution of Lyman break galaxies), 
the {\em tangential} distance distribution of quasars is investigated.

(iii) However, two-dimensional wedge diagrams are also plotted to show
clear qualitative evidence of the tracing of large scale structure by
the quasar distribution.

Details of the method and selection of homogeneous quasar 
samples are described in \SS\ref{s-method}.
Results are presented in \SS\ref{s-results}, 
and discussion and conclusions are presented
in \SS\ref{s-conc}.

A perturbed Friedmann-Lema\^{\i}tre-Robertson-Walker cosmological model
is assumed here. The context in which quasars could reasonably be
expected to form a tracer population of large scale structure may
be any model in which quasars form in galaxy centres or in which 
galaxies form around quasars. The expected short life times of quasars 
should not prevent them from forming a tracer population, though they
might form a biased population which could either weaken or strengthen
the amplitude of the signal.

The Hubble constant is parametrised here 
as $h\equiv H_0/100\,$km~s$^{-1}$~Mpc$^{-1}.$ Comoving coordinates are
used throughout 
[i.e. `proper distances', eq.~(14.2.21), \citet{Wein72}, 
equivalent to `conformal time'
if $c=1$].
Values of the density parameter, $\Omega_0$, and 
the dimensionless cosmological constant,
$\lambda_0$, are indicated where used.


\fzhist

\fallsky

\section{Method} \label{s-method}

\subsection{Choice of catalogue and sky regions} \label{s-choosecat}

There are now around $10^4$  quasars  which have 
publicly available redshifts and celestial positions \citep{Veron98}.
What is the optimal way to search for the power spectrum peak among
these data? 

Figs~\ref{f-zhist} and \ref{f-allsky} show the redshift
and sky distributions of these quasars. Depending on the 
redshift range of any quasar sample, a few hundred {\hMpc} 
typically correspond to $\Delta z \sim 10^{-1}$ in redshift to within
an order of magnitude, and at $z \sim 2$, where the peak in the
redshift distribution lies, to $\Delta \theta \sim 1 \deg.$ 

There is clearly structure in the combined redshift histogram
for the full sky at around this scale. As discussed in 
detail by \citet{Scott91}, the ratios in wavelength 
of important emission lines which contribute
to the chance of detecting a quasar (Ly$\alpha$, \CIV, \CIII, \MgII)
correspond to intervals 
$\Delta \ln \lambda \approx 0.2$ and clearly contribute 
to the obvious peaks in the distribution. 

The alternative to searching for structure in the redshift 
direction is to search in the tangential direction.
Most of the deep surveys visible in Fig.~\ref{f-allsky} are
based on photometric selection from photographic plates of size roughly 
$6\deg \times 6\deg,$ in particular from objective prism surveys.
This is moderately larger than the scale of interest. 

In order to minimise possible systematic effects due to obscuration
by dust and modification of the sky background by bright stars, 
surveys near the South or North galactic poles would be best.

Near the South galactic pole (SGP), 
several regions have been observed contiguously,
with (at least to the eye) a reasonable homogeneity across the 
different plates in the region.

\fzVeron

\fskyV

\tfourier

\fcone

More objectively, the largest single homogeneous subset
of the \citet{Veron98} catalogue in the
SGP region is that of \citet*{IovCS96}, who used an
`automatic quasar detection' method, i.e. applied a computer 
algorithm to the digitised images of objective prism plates
taken on the UK Schmidt Telescope at Siding Spring, Australia.

This is the catalogue chosen for analysis here. 
The overall redshift distribution of this catalogue is 
shown in Fig.~\ref{f-zVeron}, and the sky distribution of 
the $1.8 \le z < 2.4$ component is shown in Fig.~\ref{f-skyV}.

Wedge diagrams of the catalogue, 
within the limits of Table~\ref{t-fourier} (see \SS\ref{s-method1})
are shown in Fig.~\ref{f-cone}. 
Note that the right ascension and declination subsamples are not
entirely independent sets of quasars (see Fig.~\ref{f-skyV}),
though since only one angular coordinate is used in each analysis,
they are very close to being effectively independent.
Although quantitative 
analysis in these planes would be difficult to carry out 
due to the obvious 
redshift selection effects, it is qualitatively clear that
a void-like structure is present, at a scale
near $\llss\sim 130${\hMpc} for $(\Omega_0=0.3, \lambda_0=0.7)$.
In order to avoid the redshift selection effects 
(see also \citealt{HarSch90,Scott91}), the analysis here is restricted
to Fourier space.

\subsection{One-dimensional Fourier Analysis} \label{s-method1}

The study of structure at $\llss$ is made
by one-dimensional 
Fourier analysis of two subsamples of the \citet{IovCS96}
sample 
[i.e. of the subset
of the \citeauthor{Veron98} 
catalogue for which redshifts are obtained from
\citet{IovCS96}]
which have maximum tangential survey length. That is,
one subsample combining three plates in the right ascension 
direction and one subsample combining three plates in the
declination direction are chosen. 

The right ascension and declination 
boundaries are chosen conservatively, i.e. to within at 
most 2.5 great circle degrees from the plate centres, 
and are shifted even closer in where it looks like there
may be incompleteness close to the boundary. Although what
appears to be a lack of quasars near a few plate boundaries might
in fact be due to real voids, it is preferable to risk losing
some real signal rather than risk including some noise.

The angular limits chosen are indicated in Table~\ref{t-fourier}.
The redshift range used is $z_1 = 1.8 \le z < z_2 = 2.4$, which
includes most of the catalogue, and is small enough 
to superimpose only a few
`units' of large scale structure, i.e. the
signal should not be significantly reduced 
by the  superimposition of structures which
are out of phase.

In each of the two subsamples, the angular positions in the
long direction (right ascension and declination respectively)
are converted to comoving tangential lengths $\dperp$ by 
\begin{eqnarray}
\dperp (z,\theta) &\equiv& 
(\theta - \theta_1) \; \dpm(z) \nonumber \\
&=& (\theta - \theta_1) 
	\left\{ 
        \begin{array}{lll}
        R_C \sinh [d(z)/R_C] , & \kappa_0 < 0 \\
        d(z) , & \kappa_0 = 0 \\
        R_C \sin [d(z)/R_C] , & \kappa_0 > 0.
        \end{array}
        \right. ,
\label{e-defpm}
\end{eqnarray}
where
the angular position 
$\theta=\alpha \cos \delta $ or $\theta=\delta$ and the survey limit is
$\theta_1=\alpha_1 \cos\delta_1$ 
or $\theta_1=\delta_1$ respectively (in radians), 
$\dpm(z)$ is the proper motion distance,  
\begin{equation}
d(z) 
= {c \over H_0} \int_{1/(1+z)}^1 
{ \mbox{\rm d}a \over a \sqrt{\Omega_0 /a - \kappa_0 + \lambda_0 a^2} },
\label{e-defdprop}
\end{equation}
is the proper distance 
[eq.~(14.2.21), \citet{Wein72}], 
\begin{equation}
\kappa_0 \equiv \Omega_0 + \lambda_0 -1
\end{equation}
is the (dimensionless) curvature of the observational Universe and 
\begin{equation}
R_C \equiv {c \over H_0} { 1 \over \sqrt{ | \kappa_0| } }
\end{equation}
is its curvature radius.

The maximum tangential length considered is
$\dperp(z_1,\theta_2)$, where $z_1=1.8$ is the low redshift limit
of the sample. This is the minimum tangential length corresponding
to $\theta_2-\theta_1$ over the range in $z$ and 
the $(\Omega_0, \lambda_0)$ domain considered in this paper.
For negative or zero curvature, $\dperp$ is always an increasing
function of $z$.
For positive curvature and $\Omega_0 < 1.1, \lambda_0 < 1, $ the 
$d/R_C = \pi/2$ point 
(halfway to the antipode)
occurs at $ z \gtapprox 8 ,$ so the domain of decreasing $\dperp$ 
is not reached here. Thus, $\dperp$ 
is a strictly increasing function of $z$ in the domain of interest 
of this paper and $\dperp(z_1)$ provides the minimum tangential 
length.

Note that the choice of this cutoff throws away a small amount of 
data (e.g. for which $z \approx z_2, \theta \approx \theta_2$),
but ensures that the one-dimensional number density 
distribution d$N/$d$\dperp$ is a uniform projection of a (large) 
subset of the
two-dimensional distribution d$^2N/$d$\dperp$d$\dpm$.
Inclusion of the small amount of 
lost data would create a nonuniform number density
projection and would introduce non-physical 
power to the Fourier transform, so is not attempted here.

For values of $(\Omega_0, \lambda_0)$
in the range $0.0 \le \Omega_0 \le 1.1, -0.1 \le \lambda_0 \le 1.0$,
the list of positions $0 \le \dperp\le \dperp(z_1,\theta_2)$
is binned into 1024 bins and fast Fourier transformed to 
a function $f(\nu |\Omega_0,\lambda_0)$. The results below are
found to change slightly but insignificantly if fewer bins, 
e.g. 128 bins, are used. The contours in the $(\Omega_0,\lambda_0)$
plane are more noisy with fewer bins.

Two null hypotheses are considered here for the Fourier spectrum for 
each 
pair $(\Omega_0, \lambda_0)$: the possibility that large scale structure
is undetectable in the two quasar subsamples, and the possibility that
the best estimate of the scale of 
large scale structure is $\llss \pm \Delta \llss$.

\subsection{Null Hypothesis ${\cal H}_1$: No LSS peak is detectable}
\label{s-hyp1}

Informally, ${\cal H}_1(\Omega_0,\lambda_0)$ is the hypothesis that there is no
peak in the tangential distributions of the quasars 
due to large scale structure.

More precisely, ${\cal H}_1(\Omega_0,\lambda_0)$ is the hypothesis that:
\begin{list}{(\roman{enumi})}{\usecounter{enumi}}
\item  the pair $(\Omega_0,\lambda_0)$ is correct;
\item 
the value of the Fourier transform $f$ at the large scale structure 
frequency $\nu_0 \equiv 1/\llss$, i.e.
$f(\nu_0 |\Omega_0,\lambda_0)$,
 is not significantly higher than that
expected from the 
distribution of 
the same statistic evaluated for Poisson distributions in $\theta$, 
for  fixed values of $\nu_0, \Omega_0,\lambda_0$,
for both subsamples;
and
\item
$\numax$, defined as the local maximum in $f(\nu |\Omega_0,\lambda_0)$ 
at the greatest value of $\nu$ satisfying $\nu < 0.01${\hMpcinv},
is not significantly higher than that
expected from the distribution of 
the same statistic evaluated for Poisson distributions in $\theta$,
for  fixed values of $\Omega_0,\lambda_0$,
for both subsamples.
\end{list}

The Poisson distributions are pseudo-random samplings of uniform
distributions in $\theta$ within each of the three subdivisions
of the sample as defined as in Table~\ref{t-fourier}. 
The number of simulations calculated is 30.
A Gaussian 
smoothing of standard deviation two bin widths is applied 
to the Fourier transform $f(\nu|\Omega_0,\lambda_0)$
before
searching for the local maximum.

The purpose of criterion (ii) is that 
if a peak is present at the scale expected, then this may contribute
to rejecting ${\cal H}_1$ by the presence of a strong peak.
However, 
because the overdensity may not be very high, this may not be sufficient
in itself to reject ${\cal H}_1$. 

An independent and possibly more sensitive test is (iii): is the
best estimate of the frequency of a peak significantly different 
from that for Poisson distributions, independently of any criterion
on the absolute height of the peak?
Likely values of $\left<\numax \right>$ for Poisson 
simulations are around $0.005${\hMpcinv},
though this depends on the smoothness or roughness of $f$ for
the Poisson simulations.
Since $1/\llss \approx 0.0077${\hMpcinv}, then, as long as the scatter
in $\numax$ for the simulations is small enough, criterion (iii) 
may enable rejection of ${\cal H}_1$ for pairs $(\Omega_0,\lambda_0)$
which correctly describe the observational Universe, because in that
case (under the principle assumed for this paper),
the peak is expected to occur at a special frequency rather than at
an arbitrary frequency.

\subsection{Null Hypothesis ${\cal H}_2$: 
The best estimate of the frequency of an LSS peak is at $1/\llss$}
\label{s-hyp2}

The more interesting hypothesis is 
${\cal H}_2(\Omega_0,\lambda_0)$, the hypothesis that the best estimate
of a peak in the tangential distributions of the quasars 
due to large scale structure is at $\llss \pm \Delta \llss,$
independently of whether the peak is significant or not. 

${\cal H}_2(\Omega_0,\lambda_0)$ is quantified as follows:
\begin{list}{(\roman{enumi})}{\usecounter{enumi}}
\item the pair $(\Omega_0,\lambda_0)$ is correct;
\item $\numax$ (defined as for ${\cal H}_1$) is consistent with
$\numax=1/\llss$ where the $1\,\sigma$ uncertainty in the external estimate
of $\llss$ is $\Delta \llss$, and $\Delta \numax$, 
the $1\,\sigma$ uncertainty in estimating 
$\numax$ from the present data is obtained robustly by 
bootstraps (e.g. \citealt*{BSB84}).
\end{list}
 
The estimate $\Delta \llss = 10${\hMpc} is adopted here.

\fhyponera
\fhyponedec

\fhyponeall

The bootstrap method (e.g. \citealt{BSB84}) for a catalogue of 
$N$ objects, is for $N$ objects to be randomly drawn from the same sample,
{\em allowing multiple sampling} of single objects. The statistical
uncertainties in the properties of interest
are then estimated by running several such
bootstrap simulations, which are considered as independent 
experiments. This provides an upper estimate to the uncertainty. 
The number of bootstraps used here is 30.

The bootstrap $1\,\sigma$ uncertainty and $\Delta \llss$ are assumed
to be independent and to arise from Gaussian distributions, so
are combined in quadrature.

Although ${\cal H}_2$ could in principle be consistent with the
data for $(\Omega_0,\lambda_0)$ pairs which are also consistent with
${\cal H}_1,$ the $(\Omega_0,\lambda_0)$ pairs for which 
${\cal H}_1$ is rejected and 
${\cal H}_2$ is not rejected are obviously of most interest.

\section{Results} \label{s-results}

\subsection{${\cal H}_1$: Can the absence of a peak be rejected?}
\label{s-reshyp1}

Confidence levels 
$1-P_\alpha[f(\nu_0)],$ 
$1-P_\alpha(\numax),$ 
$1-P_\delta[f(\nu_0)]$ and
$1-P_\delta(\numax),$ 
for rejecting ${\cal H}_1$ are
shown in Figs~\ref{f-hyp1ra_} and \ref{f-hyp1dec} for the
right ascension and declination subsamples respectively.

These are defined by the probability of the observational 
results given the hypothesis:
\begin{eqnarray}
P(t) \equiv \pgauss(t,\overline{t},\Delta t) &\equiv&  
\int_{|t - \overline{t}|\, / \Delta t}^\infty 
{1 \over \sqrt{2\pi}} e^{-u^2/2} \; du, \nonumber \\
&=& {1 \over 2} 
\mbox{\rm erfc}
\left( {1 \over \sqrt{2}} { |t - \overline{t}|\over \Delta t}  \right) 
\label{e-pgauss}
\end{eqnarray}
where $t$ is the parameter studied 
(either $f(\nu_0)$ or $\numax$), and $\overline{t}$ 
and $\Delta t$ are
the mean value of $t$ 
and the standard deviation of $t$ 
obtained from the Poisson simulations.

Since the question of interest is to find $(\Omega_0,\lambda_0)$
pairs for which there is, ideally, an excess of power at $\llss$ and
a frequency higher than that for Poisson distributions
(given the limits for searching for a local maximum defined 
in \SS\ref{s-hyp1}), the probability above is defined to be one-sided.

The upper panel of 
Fig~\ref{f-hyp1ra_} shows that if there is a peak at $\llss$ in 
the right ascension subsample, then it is not strong enough to 
significantly reject
the null hypothesis ${\cal H}_1(\Omega_0,\lambda_0)$ 
of the non-existence of a peak for any pair 
$(\Omega_0,\lambda_0).$ Note, of course, that 
non-rejection of ${\cal H}_1$ does not imply that ${\cal H}_1$
is correct. 
It just implies (states) that ${\cal H}_1$ is not rejected.

However, the lower panel of 
Fig~\ref{f-hyp1ra_} shows that the best estimate of the frequency
of a peak, independently of its significance, {\em is} rejected 
at the $1-P>84\%$ level for a large (though noisy) band in the 
$(\Omega_0,\lambda_0)$ plane, for $\Omega_0 \sim 0.4\pm0.1.$

Both the $f(\nu_0)$ amplitude test and the 
$\numax$ test for the declination subsample 
(Fig~\ref{f-hyp1dec})
independently confirm, to $\sim \pm 0.2$ precision in 
$\Omega_0$ and $\lambda_0$, the region of the 
$(\Omega_0,\lambda_0)$ plane which enables $1\,\sigma$ rejection of 
${\cal H}_1(\Omega_0,\lambda_0)$. A small band within the latter,
for the $\numax$ test, enables rejection to $2\,\sigma,$ 
i.e. $1-P > 98\%$.

As noted above,
the fact that the $f(\nu_0)$ test for the right ascension sample 
does not reject ${\cal H}_1$ is not a problem for alternative
hypotheses to ${\cal H}_1$ (in particular for 
${\cal H}_2$). The failure of one test to reject a model does not 
imply that it is correct, and in the presence of independent tests
which {\em do} reject the model, the overall result should be to 
reject the model. This is expressed mathematically 
as follows.

The results for the two tests, for the two subsamples, imply a
confidence level 
\begin{equation}
1 - P = 
1 - \;
P_\alpha[f(\nu_0)] \;\;
P_\alpha[\numax]\;\;
P_\delta[f(\nu_0)]\;\;
P_\delta[\numax]
\label{e-confall} 
\end{equation}
for rejecting ${\cal H}_1.$ 
Note that although the $f(\nu_0)$ and $\numax$ tests
seem to be independent (e.g. Figs~\ref{f-hyp1ra_}, 
\ref{f-hyp1dec}), this 
has not been strictly proven, so the combined confidence levels
for rejecting ${\cal H}_1$ may be slightly overestimated.

The final 
contours in confidence levels for rejecting 
${\cal H}_1(\Omega_0,\lambda_0)$
are shown in Fig.~\ref{f-hyp1all}.

These show that the null hypothesis of the absence of the large scale
structure peak is rejected at at least the $1-P > 84\%$ level 
for nearly all pairs $(\Omega_0,\lambda_0)$ with 
$0 \ltapprox \Omega_0 \ltapprox 0.6$ and at the 
 $1-P > 98\%$ level for $0.1 \ltapprox \Omega_0 \ltapprox 0.5$.
Moreover, for a band running from 
$(\Omega_0\approx 0.15,\lambda_0=-0.1)$  to
$(\Omega_0\approx 0.4,\lambda_0=1)$, 
${\cal H}_1(\Omega_0,\lambda_0)$ is rejected at the 
$1-P > 99.9\%$ level.

In other words, if the matter density of the Universe is low,
then the possibility that there is no large scale structure peak 
at $\llss=130${\hMpc} in the quasar sample is rejected, and
it is rejected to high significance for the most favoured
values of the $(\Omega_0,\lambda_0)$ pair: a low density hyperbolic
model with $(\Omega_0\approx 0.2,\lambda_0=0)$, 
or a low density flat model with 
$(\Omega_0\approx 0.3,\lambda_0=1-\Omega_0)$.

For a flat, critical density model, $(\Omega_0=1,\lambda_0=0)$,
${\cal H}_1$ is not rejected. Could it be argued that if 
$(\Omega_0=1,\lambda_0=0)$ is correct, then the 
rejection of ${\cal H}_1$ for low values of $\Omega_0$ is simply
an artefact due to making a wrong assumption? 

A simple, quantified counterargument to this 
is the following. If a peak really is present in the data, even though
its amplitude may be low, then even for incorrect values of 
$(\Omega_0,\lambda_0)$ it is likely that non-random frequencies 
can be detected. Since the search for the frequency starts just a little
above $1/\llss$, then if a low value of $\Omega_0$ is correct, the
case of 
$(\Omega_0=1,\lambda_0=0)$ may lead to the detection of a low 
frequency harmonic.
This is indeed the case. 
By substituting the word `lower'
for `higher' in (iii) of the definition of 
${\cal H}_1$ (\SS\ref{s-hyp1}), and recalculating the
equivalent of Fig.~\ref{f-hyp1all},
the confidence level for 
rejecting ${\cal H}_1(\Omega_0=1,\lambda_0=0)$ is found to be at
the $1-P > 98\%$ level.

\subsection{${\cal H}_2$: What pairs $(\Omega_0,\lambda_0)$ 
are consistent with the frequency of the peak being at $1/\llss$?}
\label{s-reshyp2}

\fhyptwora
\fhyptwodec

\fhyptwoall

Given that ${\cal H}_1$ is strongly rejected for interesting pairs 
of $(\Omega_0,\lambda_0)$ values, what are the pairs
$(\Omega_0,\lambda_0)$ which are consistent 
with $\llss=130\pm10${\hMpc}? 

For consistency, this question is formally answered by trying to 
reject the null hypothesis ${\cal H}_2(\Omega_0,\lambda_0)$,
according to which 
the frequency of the peak is assumed to be at 
$\llss=130\pm10${\hMpc}. Two-sided confidence intervals 
are used
[a factor
of two is inserted in front of the integral in Eq.~(\ref{e-pgauss})],
since the question of interest is now how close the best estimate
of the frequency is to the hypothesised frequency, and both 
low and high frequencies would reject the hypothesis.

The region in the 
$(\Omega_0,\lambda_0)$ plane for which ${\cal H}_2$ 
cannot be significantly rejected provides
an estimate of the values of $\Omega_0$ and $\lambda_0$.
In the present case, as for most other methods which provide 
significant constraints on the 
curvature parameters, there is a degeneracy between the latter, 
so that 
an estimate can only be provided for the relation between them,
rather than for both parameters independently.
A linear fit is used to describe this relation.

Figs~\ref{f-hyp2ra_} and ~\ref{f-hyp2dec} show the confidence 
levels for rejecting  ${\cal H}_2(\Omega_0,\lambda_0)$ 
for the right ascension and declination subsamples respectively.
Neither subsample is sufficient on its own to provide a precise
estimate of an $(\Omega_0,\lambda_0)$ relation, although both
reject $\Omega_0 \gtapprox 0.5$ to $\gtapprox 70\%$ significance.

However, the fact that they are independent subsamples implies that
they can be combined via
\begin{equation}
1 - P = 
1 - \;
P_\alpha[\numax]\;\;
P_\delta[\numax]
\label{e-confhyp2} 
\end{equation}
to give Fig.~\ref{f-hyp2all}, which shows
that most values of 
$(\Omega_0,\lambda_0)$ can be rejected to $> 68\%$ confidence
if $|\Omega_0 - 0.25| \gtapprox 0.15$.

More precisely, 
a linear relation between the two curvature parameters can be 
fitted to the points 
in Fig.~\ref{f-hyp2all} 
for which the probability $P$ of 
obtaining the observations is highest, using the
$68\%$ confidence limits as uncertainties,  
by linear regression of $\Omega_0$
as a function of $\lambda_0$.

This relation is 
\begin{equation}
\Omega_0 =  (0.24\pm0.04) + (0.10\pm0.08)\,\lambda_0. 
\label{e-linear}
\end{equation}
The uncertainties here relate to the fitting procedure. Since the
$1\,\sigma$ (68\% confidence) 
uncertainties on $\Omega_0$ for each value of $\lambda_0$
are not mutually independent, the uncertainties in Eq.~\ref{e-linear}
do not represent measurement uncertainties.

Since the measurement uncertainty (including $\Delta \llss$) 
is already expressed in the contours in Fig.~\ref{f-hyp2all},
this is restored to the zero-point of the relation, giving
\begin{equation}
\Omega_0 =(0.24\pm0.15)   + (0.10\pm0.08)\,\lambda_0,
\label{e-linear2}
\end{equation}
where the zero-point uncertainty includes known measurement
uncertainties and the slope uncertainty relates to the fitting
procedure.

Note that, as revealed by this relation, the point least rejected 
by the data for $\lambda_0 \equiv 0$ is much closer to the large
$\Omega_0$ $68\%$ confidence limit than to the 
low $\Omega_0$ $68\%$ confidence limit,
so the representation by Gaussian uncertainties is 
not an optimal approximation. However, if one deduces 
$\Omega_0=0.24\pm 0.15$ for the case $\lambda_0\equiv 0$ and assumes
that this is a Gaussian $1\,\sigma$ uncertainty, then this will
be sufficient for most applications, where one prefers
overestimates of uncertainties to underestimates.

Alternatively, the $\lambda_0 \equiv 0$ result can 
be written as $\Omega_0=0.24^{+0.05}_{-0.15}$.

For a flat universe, i.e. $\lambda_0 \equiv 1- \Omega_0,$ 
use of the uncertainties in Eq.~(\ref{e-linear2}) as Gaussian
uncertainties and combination in quadrature yields
$\Omega_0=0.30\pm0.15$, which is consistent with 
Fig.~\ref{f-hyp2all}.

A large part of the uncertainty here is due to the bootstraps.
For example, if the intrinsic measurement uncertainty due to the
bootstraps is removed, and only the uncertainty 
$\Delta \llss=10${\hMpc} is
used for the equivalent of Fig.~\ref{f-hyp2all}, then the regions
not rejected by the confidence level contours shrink considerably,
and $\lambda_0 < 0.4 $ is rejected at the $1-P > 95\%$ level.

However, although bootstrap estimates of uncertainties
provide an upper estimate to uncertainties, i.e. the true
uncertainties may be smaller, it is prudent to retain the bootstrap
estimate.

\subsection{Amplitude of large scale structure peak}

Since the main aim of the present study is to use the $\llss$ scale
as a ruler for measuring the curvature parameters, a statistically 
robust estimate of the {\em amplitude} of the power spectrum peak
used as a standard ruler is beyond the scope of this paper. 

Indeed, as can be seen in Fig.~\ref{f-hyp1ra_} (upper plot), the
amplitude of the peak in 
the right ascension sub-sample is insufficient (on its own) to 
significantly rule out the hypothesis of no peak at all, with respect
to the Poisson simulations. Of course, the search for a local maximum
{\em does} find that this is closer to $\nu_0$ than expected randomly,
for a certain band in the $(\Omega_0,\lambda_0)$ plane. 

However, as a guide to what might be expected in future less 
sparse surveys, the amplitude in the declination subsample,
which has a stronger signal than that of the right ascension 
subsample, may be useful to quantify, though caution 
is recommended in the interpretation of this estimate.

For the pair $(\Omega_0=0.3,\lambda_0=0.7)$ (see 
\SS\ref{s-conc}), a crude estimate of the amplitude expressed as 
a signal-to-noise ratio,
\begin{equation}
A \equiv  
{ f(\nu_0| \mbox{\rm obsvn})  - \left< f(\nu_0| \mbox{\rm Poisson}) 
\right>
\over 
\left< f(\nu_0| \mbox{\rm Poisson}) \right> } 
\pm { \sigma[f(\nu_0| \mbox{\rm Poisson})] \over
\left< f(\nu_0| \mbox{\rm Poisson})\right>  }
\end{equation}
is $A \approx 1.7 \pm 0.5$.    

This value is lower than the corresponding value in 
\citeauthor{Bro90}'s (1990) one-dimensional survey, 
for which 
$A \approx 7 $ at the $\llss$ scale (from fig.~2b of that paper), 
but is similar in order of
magnitude to the density contrast values of 
$\delta \rho / \rho \gtapprox 2.5$ found in the 
Las Campanas Redshift Survey \citepf{LCRS98}. 

This is not surprising. 
Even for $(\Omega_0=0.3,\lambda_0=0.7)$, a somewhat 
lower amplitude of density
contrast can be expected at $z\sim 2$ relative to $z \sim 0$, though
this could be compensated for (or under- or over-compensated for) by 
positive biasing of the quasar distribution relative to that of galaxies,
if quasars turn on at the densest points where galaxies are most likely
to interact and/or merge.

A full analysis of the amplitude of the signal in analyses following
the present one should potentially provide a useful constraint
on models of quasar onset and lifetimes.

\subsection{Selection Effects}

The results above are strikingly consistent with the most
recent expectations
from independent observations regarding large scale structure
and the curvature parameters: the power spectrum peak is present
at $\llss=130\pm10${\hMpc} for the popular curvature pair
$(\Omega_0=0.3\pm0.15,\lambda_0=1-\Omega_0).$ 

Could this just be a coincidence due to selection effects?
In the redshift direction,
selection effects  have long led to 
surprising results, though not to expected results, from 
quasar catalogues (see \citealt{Scott91} and references therein).

The angular scale corresponding to $z\sim 2$ and 
$(\Omega_0\sim 0.3,\lambda_0=1-\Omega_0)$ is $\sim 2\deg.$
This is half an order of magnitude smaller than the size of a
UK Schmidt plate ($\approx 6\deg$). 
This is sufficiently small that the large scale
structure scale is clearly smaller than the size of the plates,
but not so much smaller that more subtle effects related to the
plate size can be trivially excluded from contributing to
the result found above.

Possible angular selection effects, instrumental and/or 
astrophysical, in objective prism quasar surveys 
include
\begin{list}{(\roman{enumi})}{\usecounter{enumi}}
\item effects due to human subjectivity of selecting `quasar-like'  
objects from the photographic plates
\item not finding quasars in regions where the sky background is noisy
or there is no signal at all
due to step wedges, large, bright galaxies or to 
bright stars and their ghosts (due to reflection from the secondary
mirror support structure)
\item not finding quasars close to other quasars/stars 
due to overlapping spectra
\item differential apparent magnitude limits due to the vignetting 
function of the telescope plus instrument geometry
\item differences in apparent magnitude limits between plates
\item differential apparent magnitude limits due to intervening dust
\item mistaking quasars for stars in low projected number density open (star) 
clusters which happen to lie in the survey region, thereby missing
quasars in those regions
\item mistaking stars for quasars, which possibly explains the excess
numbers of objects at some specific redshifts in Fig.~\ref{f-cone}.
\end{list}

Problem (i) is avoided by application of a computer algorithm to 
digital scans of the Schmidt plates. The validity of the precise
quantification and relative weighting of the various 
`quasar-like' criteria chosen by \citet{IovCS96} to detect quasars
could be debated, but since they are calculated automatically 
over the entire scanned regions, this aspect of human subjectivity 
is applied in a consistent, objective fashion across the plates.

Problems (ii), (iii) to some extent (quasar-star overlaps), 
(vi), (vii) and (viii) are likely to be minimised by the choice of a very high
galactic latitude region, i.e. the SGP.

Problem (v) is corrected for in the control simulations by 
Poisson distributing points independently 
within the boundaries of individual plates in each of the two 
sub-samples (see \SS\ref{s-method1}).

The largest obvious contaminants listed in item (ii) 
which ocur in the fields studied
here are a moderately 
bright star near $\alpha \sim 1^h, \delta \sim -30 \deg$ 
(field 411, see table~1 of \citealt{IovCS96}) and a bright galaxy 
near $\alpha \sim 0^h50^m, \delta \sim 39\deg$ 
(field 295), which occupy less than about 0.1{\sqdeg} and 0.2{\sqdeg}
respectively. Neither corresponds to what visually appears to be 
a void in Fig.~\ref{f-skyV}.

Counting the right ascension and declination analyses
separately, the fraction of the solid angle biased by these two
objects is 
$\ltapprox [(0.2) + (0.2 + 0.1)]\mbox{\rm \sqdeg}/116.4\mbox{\rm \sqdeg}
= 0.4\%$ of the total solid angle included in the borders (as 
defined in Table~\ref{t-fourier}). This is unlikely to 
be sufficient to mimic a large scale structure signal at $\sim 2\deg$
in the full data set.

\tborders

Problem (iii) for quasar-quasar overlap could in principle
cause a weakening of the real signal
and not a false signal, although the low number density of the
quasars implies this effect should probably be small. This is because
overlaps are expected to occur at around 70{\arcs} in the dispersion
direction and 3{\arcs} in the orthogonal direction. If close quasar
pairs are missed due to this effect, then the angular distribution
measured will be {\em smoother} than the intrinsic angular 
distribution, i.e. filamentary type structure would be less easy
to detect than it should be. 

Quasar-star overlap, and problems (vi), (vii) 
and (viii) should be uncorrelated
with the intrinsic quasar distribution and are more likely to weaken
any genuine signal rather than mimic an expected signal.


\subsubsection{Differential magnitude limits?}

Problem (iv) could, in principle, provide the largest systematic
error. Fig.~3 of \citet{Dawe84}
shows that from $\approx 2\deg$ to $\approx 4\deg$ from the 
centre of the UK Schmidt Telescope field, the apparent magnitude
limit can become less faint from $\sim 0.02$~mag to $\sim 0.2$~mag
respectively. That is, less quasars would be detectable by a fixed
search algorithm towards the edges and corners of the plate than
in the middle. 

Table~\ref{t-borders} lists the angular distances, 
in great circle degrees, from the boundaries and the corners of 
the plate images, assuming that the plate centres listed in 
table~1 of \citet{IovCS96} are the correct (B1950) centres of
the actual fields observed by those authors. 

The borders of the fields, as analysed here 
(Table~\ref{t-fourier}),
are mostly about $2-2.5\deg$
from the centres, so the magnitude limits should vary much less 
than 0.1~mag over most of the plates. The furthest corners from 
the centres are mostly at about $3.3-3.4\deg$ from the centres, 
so the magnitude limit should be about 0.06~mag brighter at these
corners.

Are differences in the (solid angular) quasar number densities due to 
these magnitude limit variations visible in Fig.~\ref{f-skyV}?
Possible voids in three of the $-42\deg$ corners of 
fields in this figure, and in the $(\alpha=0^h42^m, \delta=-37.5\deg)$
corner of field \#351 (the middle declination field) 
could conceivably be related to the variation in magnitude limit.

However, since the majority of
\citeauthor{IovCS96}'s quasars 
(see fig.~2 of \citealt{IovCS96}) are roughly uniformly distributed
over an interval of about one magnitude in apparent magnitude,
a change in the magnitude limit of 0.06~mag 
could at most change the quasar number density by $\sim 6\%$.

Moreover, a small fraction of the quasars in the 
\citeauthor{IovCS96} sample have $z > 2.4$ and are not studied here.
These are likely to have the faintest apparent magnitudes, so 
$\sim 6\%$ is an {\em upper} 
estimate to the expected reduction in number density at the furthest
corners.

The apparent voids visible to the eye are presumably seen as voids
because the number density is at least an order of magnitude lower
than average. This is not explicable by a magnitude limit varying
by 0.06~mag.

In addition, in at least some of the fields, the number density 
of quasars appears to be lower in `voids' near the {\em centres}
of the plates, where the exposure ought to be deepest. 

Figure~\ref{f-skyV} suggests that
the $\delta \sim -30\deg$ field might appear to have
a circularly symmetric
central concentration of quasars, apart from a cluster/filament
near $(\alpha \sim 0^h42^m, \delta \sim -31\deg)$. This is partly
because the right ascension centres of the declination fields are
offset from one another, which is why narrow limits in right ascension
were chosen. This could affect an analysis using the boundaries adopted
here if that analysis were carried out in right ascension in the
`declination' subsample. But that is not the case here: 
only the distribution
in declination of the declination sample was studied, so the
right ascension offsets cannot affect the results.

Another reason why problem (iv) is unlikely to be significant
in the present study is that, as \citet{Dawe84} points 
out, departures from circular symmetry of the magnitude limit
are below the measurement limits (0.01~mag) of his empirical
estimates, and are not expected theoretically.

The variations in quasar density {\em around} the field centres 
appear about as strong as the variations leading to voids towards
field edges or corners. For example, the void
centred around $(\alpha=0^h50^m, \delta=-37\deg)$
in field \#351 (the middle declination field) is 
adjacent to a quite dense region around
 $(\alpha=0^h55^m, \delta=-37\deg)$, which is about equidistant
from the field centre. This strong variation cannot be caused
by the circularly symmetric vignetting of the UK Schmidt Telescope.

Finally, shifting several of 
the borders defined in Table~\ref{t-fourier} by {0.5\deg} typically
modifies the slope and zeropoint of the $(\Omega_0,\lambda_0)$
relation of Eq.~(\ref{e-linear2}) by only about $\pm 0.01$. Since magnitude
limits should, statistically, 
have {\em some} effect at the corners, even if small,
this shows that uncertainty due to differential magnitude limits 
appears to be negligible relative to the basic result.

\subsubsection{Selection effects: summary}

The possible selection effects discussed above do not seem sufficient
to provide a non-cosmological explanation for the present results.
It remains that there could be unusual systematic effects not 
reported by \citet{IovCS96}, e.g. dust on patches of the photographic
emulsion, bright supernovae, very bright comet or satellite trails. 
However,
the overall similarity in the results for the right ascension and
declination samples makes it unlikely that these could explain
the principal results.

\subsection{Could the signal be confined to just a small part of
the data set?} \label{s-zslices}

Although Fig.~\ref{f-cone} provides {\em qualitative} 
evidence that the signal comes from structures spread through
the full data set, 
it could be possible that this subjective judgment is wrong.
Could it be the case that a single structure provides
most of the signal, and that very little signal 
comes from the rest of the full data
set? 

One possible approach to investigating this is 
by subdividing the full data set
into smaller `independent' subsets.
Given the sparsity of the data set, subdivision into smaller
subsets is likely to increase the noise. The $\llss$ scale has 
a primarily statistical meaning, 
and random variation in the estimate of the value of $\llss$ 
(something like the standard error in the mean)
is likely to be greater in the smaller samples than in the larger samples.


However, in order to provide an illustrative answer to the question, 
a subdivision of the two subsamples (right ascension and declination)
was carried out, into the 
four redshift intervals 
$1.8 \le z < 1.95, $
$1.95 \le z < 2.10, $
$2.10 \le z < 2.25$ and 
$2.25 \le z < 2.40.$ These provide radial intervals which are close
to the $\llss$ scale, at least for $(\Omega_0=0.3, \lambda_0=0.7)$.
Analyses across the full angular scale were carried out for
${\cal H}_1$ and ${\cal H}_2$ as before, but dividing the data into 
eight independent subsamples instead of just two.

The result is that the hypothesis ${\cal H}_1(\Omega_0,\lambda_0)$ is
rejected more strongly than before, 
so that the equivalent of Fig.~\ref{f-hyp1all}
is rejection $1-P >98\%$ everywhere in the domain of the 
$(\Omega_0,\lambda_0)$ plane studied in this paper, 
and $1-P > 99.9\%$ over most 
of this.

This shows that the presence of signals of some sort 
is stronger in the smaller subsamples than in the larger ones. 

\fhyptwotwenty
\fhyptwodfourz

However, the length scales 
at which the signals occur in individual subsamples 
are noisier in the smaller subsamples, 
so that if no allowances are made for this, 
then the final result for ${\cal H}_2$, the hypothesis
that there is a signal {\em at the expected scale $\llss$ 
in every subsample}, is that it is
also rejected, to $1-P > 99.7\%$, for the full range of 
the $(\Omega_0,\lambda_0)$ plane covered.

In order to find consistent solutions, either the 
uncertainty $\Delta \llss$ needs to be increased, in order
to allow for the increased uncertainty per subsample, or some of 
the subsamples have to be ({\em a posteriori}) dropped.

Since the number of objects per subsample decreases by roughly a factor
of four, a doubling of $\Delta \llss$ to  $\Delta \llss = 20${\hMpc}
should be sufficient to allow for consistency with
the analysis for the full redshift range. 
This results in a small region which provides
a $1-P > 95\%$ solution in a small patch close to
the $(\Omega_0=0.3, \lambda_0=0.7)$ point, but suggesting 
a hyperbolic universe model (Fig.~\ref{f-hyp2_20_}).

Alternatively, retaining   $\Delta \llss = 10${\hMpc} and just
considering the four redshift intervals of the declination subsample
leads to Fig.~\ref{f-hyp2_d4z}. This shows a band in the
$(\Omega_0,\lambda_0)$ plane within the $1-P > 68\%$ contour, 
which is clearly consistent with (though narrower than) 
the corresponding contour for
the declination sample considered as a single sample 
(Fig.~\ref{f-hyp2dec}).

\fdhist

It is therefore clear that the signal is present in most redshift
interval subsets of the full data set, but that because of the
increased noise, in order to combine those signals into a consistent
solution, a more sophisticated technique than that presented here
would be necessary.

\section{Discussion and conclusions} \label{s-conc}

It has been shown in this paper that the use of the power 
spectrum peak corresponding to large scale structure as a standard
ruler in the tangential distribution
of a homogeneous quasar survey at $z\sim 2$ provides a new and
independent method of constraining the curvature parameters.

More precisely, what appears to be 
the optimal choice of homogeneous tangential 
quasar surveys publicly available was chosen and analysed.
The quasar data analysed were the
right ascension and declination subsamples 
(Table~\ref{t-fourier})
of the \citet{IovCS96} survey as provided in the 
\citet{Veron98} quasar catalogue.

The null hypothesis ${\cal H}_1(\Omega_0,\lambda_0)$ 
according to which 
no peak can be detected at $\llss=130${\hMpc} in the one-dimensional
Fourier transform of the tangential proper motion distance distributions,
and according to which 
the best estimate of the frequency of any such peak is 
random, is rejected to high significance for most `interesting'
pairs of $(\Omega_0,\lambda_0)$ in the range
$(0.0 \le \Omega_0 \le 1.1,-0.1 \le \lambda_0 \le 1.0).$
The highest rejection (at a confidence level 
$1-P > 99.9\%$) is for a band in the $(\Omega_0,\lambda_0)$
plane running from $\sim (0.15,-0.1)$ to $\sim (0.4,1.0).$

Inversion of the frequency condition (iii) 
of ${\cal H}_1(\Omega_0,\lambda_0)$,
by replacing `higher' by `lower', and recalculating confidence levels
results in a rejection of ${\cal H}_1(\Omega_0=1,\lambda_0=0)$ at 
$1-P > 98\%.$

Since ${\cal H}_1$ is rejected, this implies that a power spectrum
peak is present in the data.

Can the power spectrum peak be seen as a `periodicity' in the
data? Fig.~\ref{f-dhist} allows the reader to judge this subjectively
for the pair $(\Omega_0=0.3,\lambda_0=0.7).$ 

Consistently with the points at 
$(\Omega_0=0.3,\lambda_0=0.7)$ 
in the upper panels of Figs~\protect\ref{f-hyp1ra_} 
and \protect\ref{f-hyp1dec}, a periodicity of significant amplitude 
should be hard to detect in the right ascension sample, but
discernable in the declination sample. This is the case 
in Fig.~\ref{f-dhist}.

Moreover, although the redshift direction is potentially plagued
by selection effects, the wedge diagrams in Fig.~\ref{f-cone} 
show clearly that large scale structure can be 
seen in the $z-\theta$ plane,  
where $\theta=\alpha \cos\delta$ or $\theta=\delta$ as above. 
Voids at a scale of around $\llss \sim 130${\hMpc} and difficult
to explain by selection effects are visible in this figure.

A best estimate for the values of $(\Omega_0,\lambda_0)$ 
{\em consistent} with the occurrence of a large scale structure
peak in the tangential quasar distribution is found 
by trying to reject
the null hypothesis ${\cal H}_2(\Omega_0,\lambda_0)$ 
according to which 
the best estimate of the frequency of a peak 
below $1/\dperp=0.01${\hMpcinv} 
in the Fourier transforms occurs at $1/(\llss\pm \Delta \llss)$.
Bootstraps from the observational data set are used to 
robustly provide an upper estimate to the 
observational uncertainty.

Based on a linear fit to the resulting estimates in the 
$(\Omega_0,\lambda_0)$ plane, the 
relation $\Omega_0 =(0.24\pm0.15)+ (0.10\pm0.08)\,\lambda_0 $ 
is considered to best summarise the constraints on 
the two curvature parameters. For zero cosmological constant 
or flat models, the curvature parameter estimates are
$(\Omega_0=0.24^{+0.05}_{-0.15},\lambda_0\equiv 0)$ and
$(\Omega_0=0.30\pm0.15,\lambda_0\equiv 1-\Omega_0)$ respectively.

These estimates of the density parameter are in remarkable agreement
with analogous estimates obtained from the kinematics of galaxy
clusters (e.g. \citealt*{CYE97}), collapsing galaxy groups
\citep{Mam93}, as well as from the baryonic fraction in clusters
\citepf{WNEF93,Mohr99} and groups \citep{HMam94}.

The estimate of the cosmological constant under the assumption of a
flat model is also remarkably close to those which now tend towards a
flat, cosmological constant dominated universe from faint galaxy
number counts, the galaxy depletion curve behind clusters, the
supernovae type Ia method \citepf{FYTY90,FortMD97,ChY97,SCP9812,HzS98}
and cosmic microwave background methods. Results from the latter are
being updated rapidly at the moment, and are dependent on numerous
assumptions which do not enter either the present method or the
supernovae type Ia method, so are not discussed here.

The present method is very independent of the 
supernovae type Ia method: in choice of astrophysical object,
in the redshift range and in the difference between using
standard candles versus standard rulers. What is the result
of combining the two?

\fhypSNe

The combination of the present result with the
relation $0.8\,\Omega_0 + 0.6\,\lambda_0 = -0.2\pm 0.1$ from 
\citet{SCP9812},
where the error is modelled as Gaussian,
results in Fig.~\ref{f-SNe}.

Although the two relations are not quite orthogonal,
they have different enough slopes that the uncertainty in 
both is considerably reduced, such that a nearly flat model
is implied {\em without using any cosmic microwave background}
information.

A linear fit to the $68\%$ confidence contour in Fig.~\ref{f-SNe},
for $\Omega_0$ as a function of $(\lambda_0 - 0.7)$, 
results in:
\begin{eqnarray}
&\Omega_0 = (0.30\pm0.02) + (0.57\pm0.11) (\lambda_0-0.7),  & \nonumber \\   
& 
 0.55 < \lambda_0 < 0.95,&
\label{e-linearSN}
\end{eqnarray}
where as before, the uncertainties from linear regression 
relate to the fitting procedure
and underestimate the true uncertainties shown in the figure.

In this case, the maximum uncertainty in $\Omega_0$ for
a given value of $\lambda_0$ in the range above is 
$\sigma(\Omega_0)=0.11.$ Restoring this to the relation as before
yields:
\begin{eqnarray}
&\Omega_0 = (0.30\pm0.11) + (0.57\pm0.11) (\lambda_0-0.7),   & \nonumber \\
&0.55 < \lambda_0 < 0.95.&
\label{e-linearSN2}
\end{eqnarray}

This combined result strongly supports the possibility that 
the observable universe satisifies a nearly flat, perturbed
Friedmann-Lema\^{\i}tre-Robertson-Walker model, where `nearly' 
is quantified as $\pm 0.1$ in the two dimensionless 
curvature parameters. It does not require prior assumptions on 
the cosmological constant, nor does it use cosmic microwave
background (CMB) data.


An independent confirmation that no systematic errors are present
in the data studied here (or else an independent 
estimate of the systematic errors) would obviously be very desirable.

If further observations confirm that large scale structure can
be used as a standard ruler, then surveys such as the 
2dF (2 degree field) 
and SDSS (Sloan Digitial Sky Survey) 
quasar surveys should provide confirmation of the
present results within the next few years.

The former risks
systematic error from the fact that  the angular scale corresponding 
to $\llss=130\pm10${\hMpc} at $z\sim 2$ is $2\deg$: this is 
the field size of the 2dF survey.
Careful correction for this, use of lower redshift quasars, and
a much higher signal-to-noise ratio would help in obtaining convincing
results from the former. 
However, since the SDSS is digital, it should
presumably avoid this problem.

The present method would be ideally suited to a deep right ascension 
survey (i.e. for a narrow band in declination but a long, e.g. 60\deg,
band in right ascension), 
which avoids any risk of dependence of any angular scale near a few
degrees but includes a fair sample of quasars at $z\sim2$, sampling
a few hundred quasars per great circle degree. 
A blind, deep, slit spectroscopic, 
right ascension survey, `blind' in the sense of taking
spectra of {\em all} objects in a very narrow declination interval,
would provide possibly 
one of the fastest ways of obtaining constraints in the
$(\Omega_0,\lambda_0)$ plane which should suffer little 
from possible selection effects. This offers a valuable cosmological
project for the planned liquid mercury LZT (Large Zenith Telescope).

Another survey to which the present method could be usefully
applied will be the combined VIRMOS (Visual and 
Infrared MultiObject Spectroscopy, on the VLT) and XMM 
(X-ray Multiple Mission, X-ray satellite) survey, which should trace
out the filaments and/or walls of several units of large scale 
structure at super-unity redshifts 
via several different astrophysical tracers: 
galaxies, quasars and hot gas.

Over the next few years,
if the precision in the present application of this method 
were increased by an order of magnitude in 
the new quasar surveys, 
then, together with more confidence in understanding
systematic errors in the supernovae type Ia method and reduction
in random errors by more detections at high redshifts, that would
imply estimates on $\Omega_0$ and on $\lambda_0$ to a 
precision of $\pm 0.01$ --- again without use of CMB data.

Interest may then shift to the other geometrical parameters:
those required to determine the size of the Universe.
As pointed out by \citet{Schw00,Schw98}, both the curvature and the
topology of space need to be known in order to know the size of the 
Universe. For background and recent reviews 
on progress in observational methods for measuring the topological
parameters, see 
\citet{LaLu95}, \citeauthor{Stark98} (1998; and following papers in that
volume), \citet{Lum98} and \citet{LR99}. Moreover, 
a side effect of a significant measurement of 
the global geometry of the Universe
would be a confirmation or refinement of the estimates of the local
geometrical parameters \citep{RL99}.

\section*{Acknowledgments}

We thank St\'ephane Colombi, Emmanuel Bertin, Guy Mathez, 
Bernard Fort, Enrique Gazta\~naga, Jasjeet Bagla, Shiv Sethi, 
Daniel Kunth and Steve Hatton and an anonymous referee 
for useful comments. Use of the
resources at the Centre de Donn\'ees astronomiques de Strasbourg 
({\em http://csdweb.u-strasbg.fr}),
the support of the Institut d'astrophysique de Paris, CNRS, 
for a visit during which part of this work was carried out,
and the support of la Soci\'et\'e de Secours des Amis des Sciences
are gratefully acknowledged. 

\subm \clearpage ::::


\begin{thebibliography}{}
\bibitem[Adelberger et al.(1998)]{Adel98} \joref{Adelberger K.~L., Steidel C.~C., Giavalisco M., Dickinson M., Pettini M., Kellogg M.}{\apj}{508}{18}{1998}

\bibitem[Barrow et {al.}(1984)Barrow, Sonoda \& Bhavsar]{BSB84} \joref{Barrow J.~D., Sonoda D.~H., Bhavsar S.~P.}{\mnras}{210}{19P}{1984}

\bibitem[Baugh \& Efstathiou(1993)]{BauE93} \joref{Baugh C.M., Efstathiou G.}{\mnras}{265}{145}{1993}
\bibitem[Baugh \& Efstathiou(1994)]{BauE94} \joref{Baugh C.M., Efstathiou G.}{\mnras}{267}{323}{1994}


\bibitem[Broadhurst(1999)]{Bro99} Broadhurst T., 1999, in `Clustering at High Redshift', ed. V.~Le~Brun, A.~Mazure, O.~Le~F\`evre, in press 

\bibitem[Broadhurst et al.(1990)]{Bro90} \joref{Broadhurst T. J., Ellis R. S., Koo D. C., Szalay A. S.}{Nature}{343}{726}{1990}
\bibitem[Broadhurst \& Jaffe(1999)]{BJ99} {Broadhurst T., Jaffe A.~H.}, {1999}, submitted \ (arXiv:astro-ph/9904348)

\bibitem[Carlberg et {al.}(1997)Carlberg, Yee \& Ellingson]{CYE97} \joref{Carlberg R.~G., Yee H.~K.~C., Ellingson E.}{\apj}{478}{462}{1997}

\bibitem[Chiba \& Yoshii(1997)]{ChY97} \joref{Chiba M., Yoshii Y.}{\apj}{489}{485}{1997}

\bibitem[da Costa(1992)]{daCosta92} da~Costa L.~N., 1992, in The Distribution of Matter in the Universe, ed. G.~A.~Mamon, D.~Gerbal (Meudon: Obs. de Paris), p163, ftp://ftp.iap.fr/ pub/from\_users/gam/PAPERS/DAECMTG/dacosta.dvi.Z

\bibitem[da Costa et {al.}(1993)]{daCosta93} da~Costa L.~N. et {al.}, 1993, in Cosmic Velocity Fields, ed. Bouchet, F., Lachi\`eze-Rey, M., (Gif-sur-Yvette, France: Editions Fronti\`eres), p475


\bibitem[Dawe(1984)]{Dawe84} Dawe J.~A., in IAU Coll. 78, Astronomy with Schmidt-type Telescopes, ed. Capaccioli~M., Dordrecht: D.~Reidel, p193


\bibitem[de Lapparent et {al.}(1986)]{deLapp86} \joref{de~Lapparent V., Geller M.~J., Huchra J.~P.}{\apj}{302}{L1}{1986}

\bibitem[Deng et {al.}(1994)Deng, Xiaoyang \& Fang]{Deng94} \joref{Deng Z., Xiaoyang X., Fang L.-Zh.}{\apj}{431}{506}{1994} 
\bibitem[Deng et {al.}(1996)Deng, Deng \& Xia]{Deng96} \joref{Deng X.-F., Deng Z.-G., Xia X.-Y.}{Chin.Astron.Astroph.}{20}{383}{1996}



\bibitem[Einasto et {al.}(1994)]{Einasto94} \joref{Einasto M., Einasto~J., Tago E., Dalton G. B., Andernach H.}{\mnras}{269}{301}{1994}

\bibitem[Einasto et {al.}(1997a)]{Einasto97corr} \joref{Einasto, J., et al.}{\mnras}{289}{801}{1997}

\bibitem[Einasto et {al.}(1997b)]{Einasto97nat} \joref{Einasto, J., et al.}{Nature}{385}{139}{1997}




\bibitem[Fort et {al.}(1997)]{FortMD97} \joref{Fort B., Mellier Y., Dantel-Fort M.}{\aap}{321}{353}{1997}

\bibitem[Fukugita et {al.}(1990)]{FYTY90} \joref{Fukugita M., Yamashita K., Takahara F., Yoshii Y.}{\apj}{361}{L1}{1990}

\bibitem[Gazta\~naga \& Baugh(1998)]{GazB98} \joref{Gazta\~naga E., Baugh C.M.}{\mnras}{294}{229}{1998}

\bibitem[Geller \& Huchra(1989)]{GH89} \joref{Geller M.~J., Huchra J.~P.}{Science}{246}{897}{1989} 
\bibitem[Giavalisco et {al.}(1998)]{GiavDCP98} \joref{Giavalisco M., Steidel C.~C., Adelberger K.~L., Dickinson M.~E., Pettini M., Kellogg M.}{\apj}{503}{543}{1998} \ (arXiv:astro-ph/9802318)


\bibitem[Guzzo(1999)]{Guzzo99} Guzzo L., 1999, in proceedings of XIX Texas Symp. Rel. Astr. \ (arXiv:astro-ph/9911115)
\bibitem[Iovino et {al.}(1996)Iovino, Clowes \& Shaver]{IovCS96} \joref{Iovino A., Clowes R., Shaver P.}{\aaps}{119}{265}{1996} 

\bibitem[Hartwick \& Schade(1990)]{HarSch90} \joref{Hartwick F.~D.~A., Schade D.}{\araa}{28}{437}{1990}

\bibitem[Henriksen \& Mamon(1994)]{HMam94} \joref{Henriksen M.~J., Mamon G.~A.}{\apj}{421}{L63}{1994}

\bibitem[Lachi\`eze-Rey \& Luminet(1995)]{LaLu95} \joref{Lachi\`eze-Rey M.,  Luminet J.-P.}{PhysRep}{254}{136}{1995} ~(arXiv:gr-qc/9605010)



\bibitem[Luminet(1999)]{Lum98} Luminet J.-P., 1999, in Concepts de l'Espace en Physique, Les Houches, 29 sep - 3 oct 1997, Acta Cosmologica, 24, (arXiv:gr-qc/9804006)

\bibitem[Luminet \& Roukema(1999)]{LR99} Luminet J.-P. \& Roukema B.~F., 1999, in Theoretical and Observational Cosmology, NATO Advanced Study Institute, Carg\`ese 1998, ed. Lachi\`eze-Rey, M., Netherlands: Kluwer,  p117 ~(arXiv:astro-ph/9901364)

\bibitem[Mamon(1993)]{Mam93} Mamon G.~A., 1993, in The N-Body problem \& Gravitational Dynamics, ed. F. Combes \& E. Athanassoula (Meudon: Obs. Paris), p188, ~(arXiv:astro-ph/9308032)

\bibitem[Mohr et {al.}(1999)Mohr, Mathiesen \& Evrard]{Mohr99} \joref{Mohr J.~J., Mathiesen B., Evrard A.~E.}{\apj}{517}{627}{1999}


\bibitem[Perlmutter et {al.}(1999)]{SCP9812} \joref{Perlmutter S. et al.}{\apj}{517}{565}{1999} ~(arXiv:astro-ph/9812133)




\bibitem[Riess et {al.}(1998)]{HzS98} \joref{Riess A.~G. et {al.}}{\aj}{116}{1009}{1998}



\bibitem[Roukema \& Luminet(1999)]{RL99} \joref{Roukema B.~F., Luminet J.-P.}{\aap}{348}{8}{1999} \ (arXiv:astro-ph/9903453)


\bibitem[Schwarzschild(1900)]{Schw00} \joref{Schwarzschild K.}{Vier.d.Astr.Gess.}{35}{337}{1900}
\bibitem[Schwarzschild(1998)]{Schw98} \joref{Schwarzschild K.}{\cqg}{15}{2539}{1998}\ [English translation of Schwarzschild (1900)]
\bibitem[Scott(1991)]{Scott91} \joref{Scott D.}{\aap}{242}{1}{1991}
\bibitem[Shaver et {al.}(1996)]{Shav96} \joref{Shaver P.~A., Wall J.~V., Kellermann K.~I., Jackson C.~A., Hawkins M.~R.~S.}{Nature}{384}{439}{1996}


\bibitem[Starkman(1998)]{Stark98} \joref{Starkman G.~D.}{\cqg}{15}{2529}{1998}

\bibitem[Tucker et {al.}(1998)Tucker, Lin \& Shectman]{LCRS98} Tucker D.~L., Lin H., Shectman S., in Wide Field Surveys in Cosmology, ed. S.~Colombi, Y.~Mellier, B.~Raban 

\bibitem[V\'eron-Cetty \& V\'eron(1998)]{Veron98} V\'eron-Cetty M.~P., V\'eron P., 1998, ESO Scientific Report 18,  http://cdsweb.u-strasbg.fr/viz-bin/VizieR?-source=VII/207/


\bibitem[Weinberg(1972)]{Wein72} Weinberg S., 1972, Gravitation and Cosmology, New York, U.S.A.: Wiley

\bibitem[White et {al.}(1993)]{WNEF93} \joref{White S.~D.~M., Navarro J.~S., Evrard A.~E., Frenk C.~S.}{Nature}{366}{429}{1993}


\end{thebibliography}
\end{document}